# Materials Requirements of High-Speed and Low-Power Spin-Orbit-Torque Magnetic Random-Access Memory

Xiang Li, *Student Member, IEEE*, Shy-Jay Lin, Mahendra DC, Yu-Ching Liao, Chengyang Yao, Azad Naeemi, *Senior Member, IEEE*, Wilman Tsai, Shan X. Wang, *Fellow, IEEE*

*Abstract*—As spin-orbit-torque magnetic random-access memory (SOT-MRAM) is gathering great interest as the next-generation low-power and high-speed on-chip cache memory applications, it is critical to analyze the magnetic tunnel junction (MTJ) properties needed to achieve sub-ns, and ~fJ write operation when integrated with CMOS access transistors. In this paper, a 2T-1MTJ cell-level modeling framework for in-plane type Y SOT-MRAM suggests that high spin Hall conductivity and moderate SOT material sheet resistance are preferred. We benchmark write energy and speed performances of type Y SOT cells based on various SOT materials experimentally reported in the literature, including heavy metals, topological insulators and semimetals. We then carry out detailed benchmarking of SOT material Pt, $\beta$-W, and $Bi_xSe_{(1-x)}$ with different thickness and resistivity. We further discuss how our 2T-1MTJ model can be expanded to analyze other variations of SOT-MRAM, including perpendicular (type Z) and type X SOT-MRAM, two-terminal SOT-MRAM, as well as spin-transfer-torque (STT) and voltage-controlled magnetic anisotropy (VCMA)-assisted SOT-MRAM. This work will provide essential guidelines for SOT-MRAM materials, devices, and circuits research in the future.

*Index Terms*—magnetic tunnel junction, spin Hall effect, spin-orbit-torque, 2T-1MTJ

## I. INTRODUCTION

Spin-orbit-torque magnetic random-access memory (SOT-MRAM) is a promising candidate to achieve faster and more energy-efficient read and write operation compared with the current in-production spin-transfer-torque MRAM (STT-MRAM). Besides, SOT-MRAM promises higher endurance and less read disturbance than STT-MRAM, which gives it great potential as an on-chip cache replacing SRAM.

Experimentally, the most advanced STT-MTJ now can achieve 1 ns switching with a current density of 10-20 MA/cm$^2$ in sub-50nm perpendicular magnetic tunnel junctions (MTJs). [1, 2] While the best SOT-MTJ at the research front now can already achieve 0.5 ns switching with a current density of 10-20 MA/cm$^2$ in sub-500nm in-plane MTJs. [3-5] When scaled-down, in-plane SOT-MTJ can potentially achieve sub-ns, and ~fJ write operation. Though the in-plane SOT-MRAM suffers from a rather large cell size due to the aspect ratio needed to reach thermal stability and 3-terminal configuration required to perform separate write and read functions, [6] the 6T-SRAM cell size which is scaling down more slowly at advanced nodes is still much larger than the in-plane SOT-MRAM cell.[7]

Recently, abundant materials research has demonstrated SOT material that can generate close to and larger than one charge-to-spin conversion efficiency ($\xi_{ST}$). Among the various SOT materials, a heavy metal such as Pt and W[8, 9], and topological materials such as $Bi_2Se_3$, $Bi_{0.9}Sb_{0.1}$, and $WTe_2$ are of particular interest.[10-12] Nevertheless, no work has integrated these materials into a practical CMOS transistor-MTJ cell and studied the write energy-delay performance. Thus, the write power and energy benchmarking conducted thus far [13-15] remain rather qualitative, ignoring the impact from the access transistor such as limited current drive and finite transistor resistance, as well as the interdependence of write energy and write speed especially at sub-ns timescale. A better understanding of this interdependence is critical for the overall optimization of cell-level SOT-MRAM performance.

In this paper, we first introduce the framework used to model the 2T-1MTJ cell, SOT-induced switching current, and current distribution along the write path. Then, using a simplified version of this framework, we show that the write energy-delay performance of the SOT-MRAM cell depends on two critical properties of the SOT layer, i.e., the spin Hall conductivity, and sheet resistance. Based on this simplified analysis, we propose guidelines to achieve sub-ns and ~fJ operation of SOT-MTJ. Next, we benchmark the write current and energy performance of various SOT materials reported in

Manuscript received December 13, 2019. The authors thank NSF Center for Energy Efficient Electronics Science (E3S) and TSMC for financial support. This research was supported in part by ASCENT, one of six centers in JUMP, a Semiconductor Research Corporation (SRC) program sponsored by DARPA. This paper is based on a paper entitled "Materials Requirements of High-Speed and Low-Power Spin-Orbit-Torque Magnetic Random-Access Memory," presented at the 2019 IEEE S3S Conference.

Corresponding authors: X.L. and S.X.W..
X.L. and S.X.W are with Materials Science and Engineering, and Electrical Engineering, Stanford University, Stanford CA 94305 USA (e-mail: xiangsli@stanford.edu, sxwang@stanford.edu).
M.D. is with Materials Science and Engineering, Stanford University, Stanford CA 94305 USA. (e-mail: mdc2019@stanford.edu). C.Y. is with Electrical Engineering, Stanford University, Stanford CA 94305 USA. (e-mail: bryceyao@stanford.edu).
S.-J.L. and W.T. are with Corporate Research, Taiwan Semiconductor Manufacturing Company (TSMC), Hsinchu, Taiwan (e-mail: sylinw@tsmc.com, y_wtsai@tsmc.com).
Y.-C.L. and A.N. are with the Electrical and Computer Engineering, Georgia Institute of Technology, Atlanta, GA, 30332 USA, (e-mail: yliao48@gatech.edu, azad@gatech.edu).

the literature based on the simplified framework. Last, we utilize the framework to explore the best thickness and resistivity of any given SOT material for the lowest write current and energy. Sputtered Pt, $\beta$-W, and $Bi_xSe_{(1-x)}$ are explored as they are compatible with industrial production.

Compared with the original paper presented at the 2019 IEEE S3S Conference[16], this paper includes a new figure (Fig. 3 and Table II) to better illustrate the switching current and energy dependence on SOT layer sheet resistance and spin Hall conductivity, as well as to benchmark the write current and energy performance of various SOT materials reported in literature. Besides, we discuss how to expand this 2T-1MTJ modeling framework to account for other variations of SOT-MRAM cell by including new cell designs, physical terms, or full-blown micromagnetic simulations (Section V).

## II. 2T-1MTJ Cell Analysis Framework

Because the significant benefits of in-plane SOT-MRAM over STT-MRAM are its lower write power and latency, we will focus on the write operation of in-plane SOT-MRAM. Also, we will concentrate on the write delay due to intrinsic magnetization switching via the SOT effect and write energy due to Ohmic loss in a 2T-1MTJ SOT-MRAM cell. Meanwhile, the array-level SOT-MRAM read and write performance can be evaluated using methodology as in references [10, 17-21].

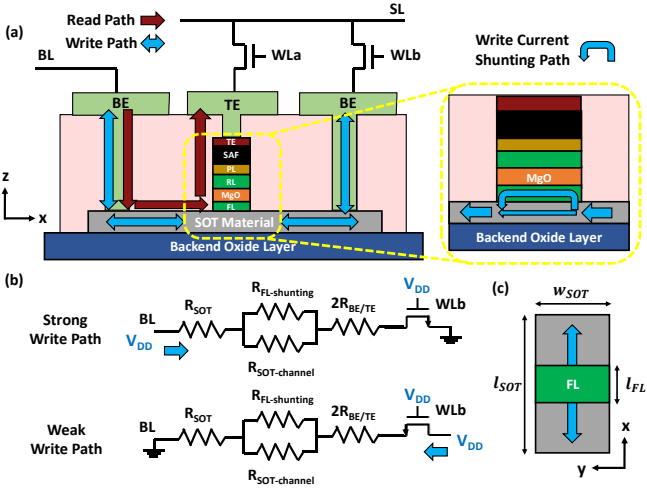

Fig. 1. (a) Cross section view of 2T-1MTJ architecture of SOT-MRAM cell. BE, TE, WL, SL, and BL refer to bottom electrode, top electrode, word line, source line, bit line respectively. FL, RL, SAF refer to free layer, reference layer, and synthetic antiferromagnetic layer respectively. (b) Equivalent circuit resistor model of the strong and weak write paths of 2T-1MTJ architecture. (c) Top view of SOT-MTJ sitting on top of the SOT material.

First, the 2T-1MTJ cell has an effective circuit model, as shown in Fig. 1(b). Here, we include the parasitic resistance from the metal electrodes, the SOT+FL layer ($R_{SOT+FL}$), and the transistor ($R_{FET}$). We also consider the write current shunting effect due to the ferromagnetic free layer (FL) sitting on top of the SOT layer, as shown in Fig. 1(a) inset. We use a square-shaped SOT-MTJ to approximate the in-plane elliptical SOT-MTJ, as shown in Fig. 1(c). The easy axis lies in the y-direction, and no external magnetic field is required to switch the magnetization. We here call this type Y SOT-MRAM. We use the PTM model of high performance 20 nm FinFET NMOS as an access transistor in Cadence Virtuoso.[22]

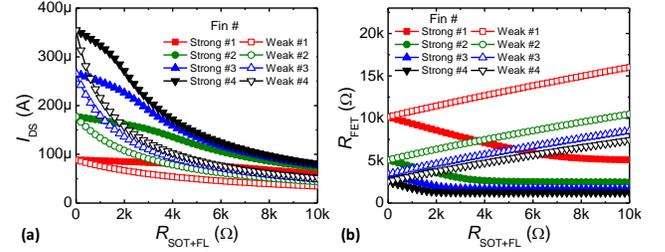

Fig. 2. N-type FinFET (a) source to drain current ($I_{DS}$), and (b) resistance as a function of $R_{SOT+FL}$ for different fin numbers. The solid (open) symbols refer to strong (weak) write path case.

Next, we model the maximal current drive when the transistor is in series with $R_{SOT+FL}$. As seen in Fig. 1(b) and Fig. 2, this will result in different current drives when the SOT current polarity flips.[23] In the strong write path case, the $V_{GS}$ value is close to $V_{DD}$, while in the weak write path case, the $V_{GS}$ is smaller than $V_{DD}$ due to the voltage drop on $R_{SOT+FL}$. Note that we use NMOS as it provides a larger current than PMOS. In later analysis, we will thus consider the weak write case only to determine how many FinFET fins are needed to perform writing.

TABLE I
2T-1MTJ SOT-MRAM CELL MODEL PARAMETERS

| Parameters | Symbol | Value | Unit |
|---|---|---|---|
| Half Metal Pitch | F | 20 | nm |
| CMOS Supply Voltage | $V_{DD}$ | 0.9 | V |
| Thermal Stability Factor of FL | $\Delta$ | 49 | |
| Damping Constant of FL | $\alpha$ | 0.01 | |
| Saturation Magnetization of FL | $M_S$ | $1\times 10^6$ | A/m |
| Shape Anisotropy Field of FL | $H_K$ | 0.168 | T |
| Effective Demag Field of FL | $\mu_0 M_{eff}$ | 0.2 | T |
| In-plane Coercivity of FL | $\mu_0 H_C$ | 0.004 | T |
| SOT Layer Dimension | $l_{SOT}, w_{SOT}$ | 100, 60 | nm |
| FL Dimension | $l_{FL}, t_{FL}$ | 20, 2 | nm |
| FL (CoFeB) Resistivity | $\rho_{FL}$ | 130 | $\mu\Omega\,cm$ |

To model the damping-like SOT-driven switching current and speed of the in-plane magnetized free layer (FL), we employ the same model used for STT-MTJ [24] with experimentally observed values as listed in Table I[4, 25]. First switching current $I_c = I_{c0}\left[1 + \ln\left(\frac{\pi}{2\theta_0}\right)/\frac{t_{sw}}{t_0}\right]$, where $\theta_0(=1/\sqrt{\Delta})$ is the initial angle of the magnetization with respect to the easy axis, $t_{sw}$ is the FL switching time, and $t_0$ is the FL characteristic relaxation time. Here, $\Delta = w_{SOT}l_{FL}t_{FL}H_K M_S/2k_B T$, $H_K = 8\pi M_S t_{FL}(w_{SOT}/l_{FL} - 1)/w_{SOT}$, and $t_0 = (1+\alpha^2)/\alpha\gamma H_K$.[26] Next, critical switching current is expressed as $I_{c0} = \frac{2e}{\hbar}\mu_0 M_S t_{FL} t_{SOT} w_{SOT}\alpha(H_C + \frac{M_{eff}}{2})/\xi_{ST}$.[20] Consistent with references [4, 25], to reduce the critical switching current, we use a rather small FL effective demagnetization field $\mu_0 M_{eff}$ of 0.2T which results from a large perpendicular anisotropy field at the CoFeB/MgO interface. We caution that the damping constant value 0.01 is taken from reference [4] and might change with the SOT material conditions. Note that the Δ value of 49 is sufficient for 128MB SRAM using error correction code (ECC).[27] We must note that to simplify the analysis, we do not consider the effects of field-like torque and the Oersted field on the magnetization switching process. We will discuss these effects in section V later.

Last, we consider the effect of shunting due to the free layer sitting on top of the SOT layer. Using a simple parallel resistor model to provide critical current $I_c$ in the SOT channel, the shunting current flowing through the FL is $I_s = \frac{\rho_{SOT} t_{FL}}{\rho_{FL} t_{SOT}} I_c = \frac{R_s^{SOT}}{R_s^{FL}} I_c$, where $R_s$ refers to sheet resistance. The write energy in the SOT+FL line is: $E_{sw}^{SOT+FL} = (I_c + I_s)^2 \frac{\rho_{SOT}(l_{SOT} - l_{FL})}{w_{SOT} t_{SOT}} t_{sw} + I_c^2 \frac{\rho_{SOT} l_{FL}}{w_{SOT} t_{SOT}} t_{sw} + I_s^2 \frac{\rho_{FL} l_{FL}}{w_{FL} t_{FL}} t_{sw}$ and the write energy in the FinFET is $E_{sw}^{FET} = I_{sw}^2 R_{FET} t_{sw}$. In practice, the write current and energy will be higher if driving the FinFET at maximum current.

### III. OPTIMIZATION OF SHEET RESISTANCE AND SPIN HALL CONDUCTIVITY

To simplify the analysis, we assume the thermal stability and cell size determine the FL properties. Then, we can write down $I_{sw} (= I_c + I_s)$ and $E_{sw}$ as a function of only two variables: apparent spin Hall conductivity $\sigma_s^* = \xi_{ST}/\rho_{SOT}$ of the SOT layer, and SOT material sheet resistance $R_s^{SOT} = \rho_{SOT}/t_{SOT}$:

$$I_{sw} \propto \left(1 + \frac{R_\square^{SOT}}{R_\square^{FL}}\right) \frac{1}{\sigma_s^* R_\square^{SOT}} = \frac{1}{\sigma_s^* R_\square^{FL}} \left(1 + \frac{R_\square^{FL}}{R_\square^{SOT}}\right);$$

$$E_{sw}^{SOT+FL} \propto \frac{1}{\sigma_s^{*2}} \left(\frac{(1+R_\square^{SOT}/R_\square^{FL})^2 a}{R_\square^{SOT}} + \frac{b}{R_\square^{SOT}} + \frac{b}{R_\square^{FL}}\right), E_{sw}^{FET} \propto I_{sw}^2.$$

where $a = \frac{l_{SOT} - l_{FL}}{w_{SOT}}$, $b = \frac{l_{FL}}{w_{SOT}}$

TABLE II
SOT MATERIAL PROPERTIES FOR SIMPLIFIED ANALYSIS
(*Growth method is sputtering unless noted otherwise, all data were obtained room temperature)

| SOT Material* | $\rho_{SOT}$ μΩ cm | $t_{SOT}$ nm | $\xi_{ST}$ | $R_\square^{SOT}$ Ω/□ | $\sigma_s$ $10^5 \frac{\hbar}{2e}(\Omega m)^{-1}$ | Ref. with Measure Method |
|---|---|---|---|---|---|---|
| Pt | 20 | 6 | 0.07 | 33 | 3.5 | [19] ST-FMR |
| Pt | 51 | 4 | 0.2 | 128 | 3.9 | [15] Harmonic |
| Pd$_{0.25}$Pt$_{0.75}$ | 58 | 4 | 0.26 | 144 | 4.5 | [13] Harmonic |
| Au$_{0.25}$Pt$_{0.75}$ | 83 | 8 | 0.35 | 104 | 4.2 | [28] Harmonic |
| [Pf/Hf]$_n$/Pt | 144 | 4.6 | 0.37 | 313 | 2.6 | [9] Switching &ST-FMR |
| β-W | 260 | 5 | 0.3 | 520 | 1.2 | [5] Harmonic |
| β-W | 238 | 5 | 0.6 | 476 | 2.6 | [31] Loop Shift |
| α-W | 20 | 7 | 0.04 | 29 | 2 | [20] Switching &ST-FMR |
| Ta | 190 | 8 | 0.12 | 238 | 0.6 | [32] Spin Hall Magneto- resistance |
| TaB | 197 | 4 | 0.2 | 493 | 1.02 | [21] ST-FMR |
| Bi$_2$Se$_3$ (MBE) | 1755 | 8 | 3.5 | 2194 | 2.0 | [29] Loop Shift |
| Bi$_2$Se$_3$ (MBE) | 1060 | 7.4 | 0.16 | 1432 | 0.15 | [10] Differenti- al Hall |
| Bi$_x$Se$_{(1-x)}$ | 2083 | 8 | 2.9 | 2604 | 1.39 | |
| | 1538 | 16 | 1.75 | 961 | 1.1 | |
| | 1428 | 40 | 0.45 | 357 | 0.32 | |
| Bi$_{0.9}$Sb$_{0.1}$ (MBE) | 400 | 10 | 52 | 400 | 130 | [12] Loop Shift |
| Bi$_{0.83}$Sb$_{0.17}$ | 1000 | 10 | 1.2 | 1000 | 1.2 | [33] Harmonic |
| Sb | 333 | 10 | 0.13 | 333 | 0.39 | |
| Bi$_{0.1}$Sb$_{0.9}$ | 375 | 80 | 0.25 | 47 | 0.66 | [34] ST-FMR |
| WTe$_2$ (Exfoliated Flakes) | 580 | 120 | 0.51 | 48 | 0.9 | [11] Harmonic &ST-FMR |
| WTe$_x$ | 435 | 5 | 0.42 | 870 | 0.97 | [30] Harmonic &ST-FMR |

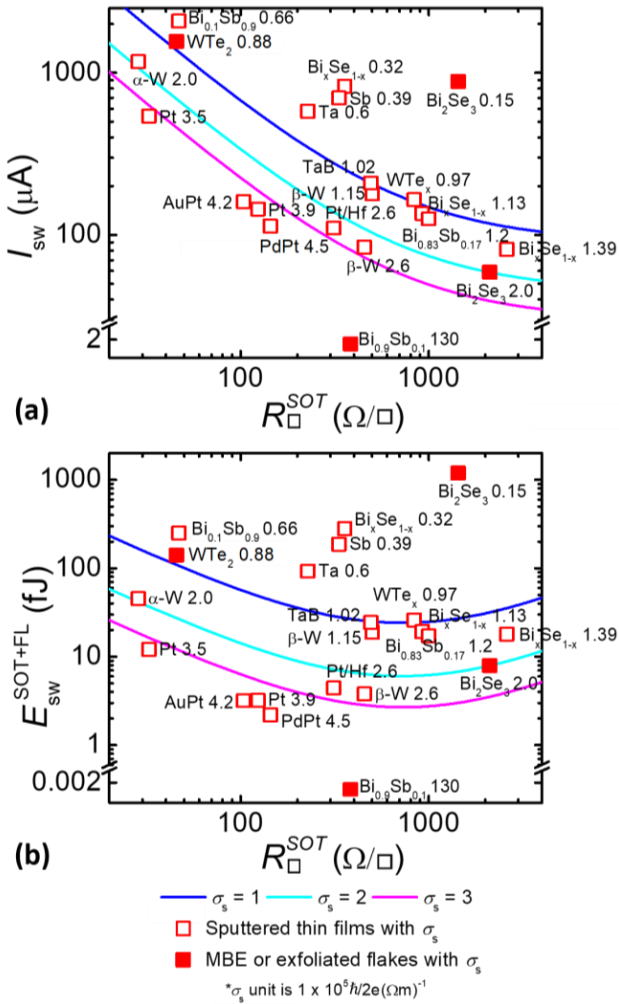

Fig. 3. (a) Switching current $I_{sw}$ and (b) switching energy in the SOT and FL $E_{sw}^{SOT+FL}$ as a function of SOT layer sheet resistance $R_\square^{SOT}$ for SOT materials with different $\sigma_s$ values. The write time is 0.5 ns. The data points are calculated based on published experimental data on various SOT materials as listed in Table II, with their $\sigma_s$ values labelled besides each data point. The open data points are taken from sputtered materials, while the solid data points from molecular beam epitaxy (MBE)-grown or exfoliated materials.

There are two scenarios here. First, when $R_{FET} \gg R_{SOT+FL}$, $E_{sw} \propto I_{sw}^2$. Hence, we only need to consider optimizing for $I_{sw}$. As seen in Fig. 3(a), large $\sigma_s^*$ and $R_\square^{SOT}$ are preferred. Taking experimental data for SOT materials in literature, as shown in Table II, we then conduct a simplified analysis using the above equations. We see that materials with too small $R_\square^{SOT}$ (< 200 Ω/□), such as Pt, Ta, and Pt-based alloys, result in a sharp increase in $I_{sw}$, thus too large transistor size. Meanwhile, too large $R_\square^{SOT}$ (> 2000 Ω/□) deteriorates the FinFET current drive, as shown in Fig. 2, thus leading to larger transistor size. Note that though the $I_{sw}$ reduces, the decrease in the current drive is more severe leading to an increase in transistor fin numbers.

Second, when $R_{FET} \ll R_{SOT+FL}$, the criteria of $I_{sw}$ optimization are similar to above. As shown in Fig. 3(b), when $R_\square^{SOT} = \sqrt{\frac{l_{SOT}}{w_{SOT}} R_\square^{FL}} \sim 1000$ Ω/□, $E_{sw}^{SOT+FL}$ is minimized. Similar to the first scenario, materials with too small $R_\square^{SOT}$ (< 200 Ω/□) result in a sharp increase in $E_{sw}^{SOT+FL}$. But because $E_{sw}^{SOT+FL}$ is inversely proportional to $\sigma_s^2$, a conductive SOT material such as Pt with very large $\sigma_s$ can still maintain a low $E_{sw}^{SOT+FL}$, as seen in Fig. 3(b).

Hence, in both scenarios, the goal is to find a SOT material with large $\sigma_s$ and moderate $R_\square^{SOT}$ of 500 - 2000 Ω/□.

It is worthwhile to discuss briefly here the different materials and experimental parameters chosen in Table II.[5, 9-11, 13, 15, 19-21] [28-34] First, the experimental $\xi_{ST}$ and $\sigma_s$ values were measured using various techniques, as shown in Table II. Hence, care must be taken when comparing the SOT-related values across different materials. Then, for each material, the $\rho_{SOT}$, $R_\square^{SOT}$, $\xi_{ST}$, and $\sigma_s$ values all change with the material thickness. Here, we list three $Bi_xSe_{(1-x)}$ thicknesses cases so as to illustrate the impact of thickness better, while we do not show the 4 and 6 nm $Bi_xSe_{(1-x)}$ cases with $R_\square^{SOT}$ larger than 4000 Ω/□. We will model the influence of $t_{SOT}$ for Pt, $\beta$-W, and sputtered $Bi_xSe_{(1-x)}$ in detail in the following section. Next, we illustrate both sputtered materials and other MBE-grown or exfoliated materials in Figure 3 and Table II. In particular, MBE-grown $Bi_{0.9}Sb_{0.1}$ shows very promising SOT performance [12] but sputtered BiSb alloys show similar performance as other heavy metals or topological materials[33, 34]. Last, it is further worth noting that there are reports on SOT of topological insulator/light metal bilayers such as $(Bi_{1-x}Sb_x)_2Te_3/Ti$, $Bi_2Te_3/Mo$, $SnTe/Ti$, and $Bi_2Se_3/Mo$. [14, 35] To simplify the analysis based on the single-layer SOT material/FL model, we do not include these bilayers in this work.

## IV. BENCHMARKING WITH SEVERAL SOT MATERIALS

Based on the above framework, we model a 2T-1MTJ cell with several representative SOT materials, including Pt, $\beta$-W, and sputtered $Bi_xSe_{(1-x)}$. Using the model parameters listed in Table I, we find the $R_{FET}$ on the range of 5 kΩ dominates over $R_{SOT+FL}$, as shown in Fig. 2 and Table III. Hence, the write energy in the transistor $E_{sw}^{FET}$ also dominates over that in the SOT and FL $E_{sw}^{SOT+FL}$.

For heavy metal, we need to consider the effect of spin diffusion length $\lambda_s$: $\xi_{ST} = \theta_{ST}[1 - \text{sech}(t_{SOT}/\lambda_s)]$, where $\lambda_s$ is the spin diffusion length of the SOT material, and $\theta_{ST}$ is the effective spin-torque efficiency when all spins diffuse into the FL adjacent to the SOT layer. From literature [5], we use a typical $\sigma_s(=\theta_{ST}/\rho_{SOT})$ and $\lambda_s$ value for $\beta$-W as listed in Table III. Note that experimentally $\rho_{SOT}$ can be tuned by sputtering conditions and remains relatively constant within a range of $t_{SOT}$ [5]. We assume that $\sigma_s$ does not change over a limited range of $\rho_{SOT}$ and $t_{SOT}$. Hence, we model $\xi_{ST}, I_{sw}, E_{sw}^{FET}, E_{sw}^{SOT+FL}$ based on two independent variables $\rho_{SOT}$ and $t_{SOT}$. As shown in Fig. 4(a-c), we obtain $\xi_{ST}$, fin number needed to provide $I_{sw}$, and $E_{sw}^{SOT+FL}$ at switching speed of 0.5 ns. As $E_{sw}^{FET} \propto I_{sw}^2$, the $E_{sw}^{FET}$ dependence on $\rho_{SOT}$ and $t_{SOT}$ is similar to the fin number.

Similarly, we use published heavy metal Pt $\sigma_s$ value and $\lambda_s \propto 1/\theta_{ST}$ relationship according to the Elliott-Yafet spin relaxation mechanism.[8] Though Pt has a higher $\sigma_s$ than $\beta$-W, its much lower $\rho_{SOT}$, thus $R_\square^{SOT}$ value results in higher $I_{sw}$ and $E_{sw}^{FET}$, while fin number of 2 can only be achieved when Pt is very resistive (relatively larger $R_\square^{SOT}$), as shown in the insets of Fig. 4(d-f) and Table III. Note that the higher $\sigma_s^*$ of Pt brings down $E_{sw}^{SOT+FL}(\propto 1/\sigma_s^{*2})$ compared with $\beta$-W.

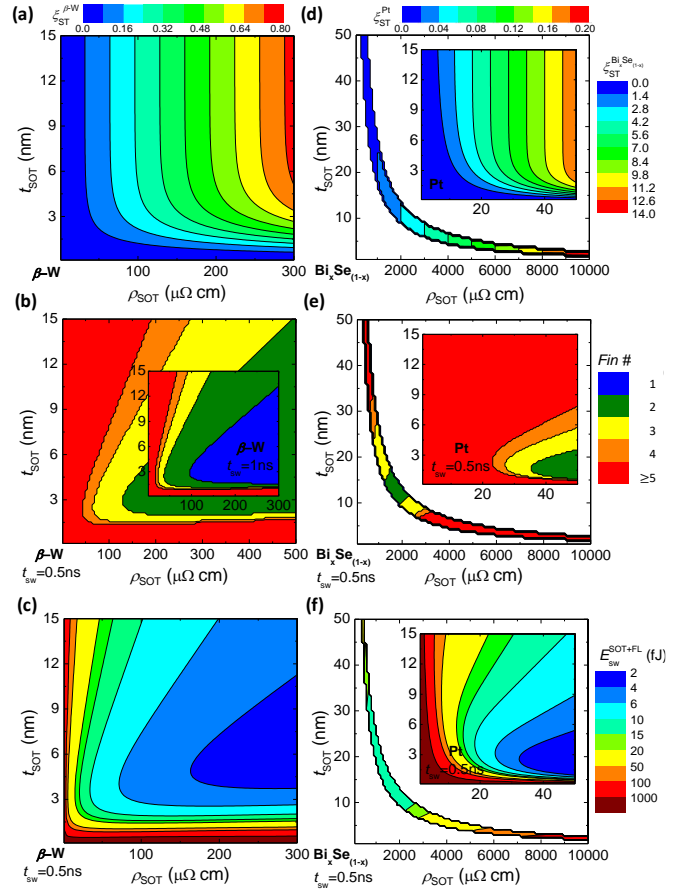

Fig. 4. Metrics as a function of $t_{SOT}$ and $\rho_{SOT}$ for $\beta$-W, $Bi_xSe_{(1-x)}$, and Pt (inset of (d-f)): (a)(d) Charge-to-spin conversion efficiency ($\xi_{ST}$); (b)(e) FinFET fin number for switching at 0.5ns (inset 1ns for (b)); (c)(f) Total switching energy for 0.5ns switching.

For sputtered Bi$_x$Se$_{(1-x)}$, spin diffusion length cannot be defined because its resistivity and band structures change drastically as a function of thickness.[10] Hence, we can only plot a certain range of $\rho_{SOT}, t_{SOT}$ as shown in Fig. 4(d-f), instead of the full 2D matrix as in the heavy metal case. As discussed above, 15 nm is preferred for low $I_{sw}$ resulting from a tradeoff between the shunting factor and $\sigma_s^*$. Its relatively low $\sigma_s^*$ results in around 2 times higher $E_{sw}^{SOT+FL}$ (10.2 fJ) than that of Pt/$\beta$-W (3.7/2.7 fJ), as shown in Table III.

TABLE III
2T-1MTJ SOT MATERIAL BENCHMARKING PARAMETERS AND RESULTS

| Parameters | Symbol | $\beta$-W | Pt | Bi$_x$Se$_{(1-x)}$ | Unit |
|---|---|---|---|---|---|
| Spin Hall conductivity | $\sigma_s$ | 2.5 | 3.5 | 1.5 | $10^5 \frac{\hbar}{2e}(\Omega m)^{-1}$ |
| Spin diffusion length | $\lambda_s$ | 1.3 | $\frac{0.1}{\theta_{ST}}$ | N/A | nm |
| Resistivity* | $\rho_{SOT}$ | 200 | 50 | 1400 | $\mu\Omega\,cm$ |
| Thickness* | $t_{SOT}$ | 5 | 3 | 15 | nm |
| SOT sheet resistance* | $R_\square^{SOT}$ | 400 | 167 | 933 | $\Omega/\square$ |
| SOT efficiency* | $\xi_{ST}$ | 0.48 | 0.17 | 1.96 | |
| Fin number* | Fin # | 2 | 2 | 2 | |
| Switching current* | $I_{sw}$ | 110 | 140 | 118 | $\mu A$ |
| SOT+FL write energy* | $E_{sw}^{SOT+FL}$ | 3.7 | 2.7 | 10.2 | fJ |
| FinFET write energy* | $E_{sw}^{FET}$ | 33.2 | 51.5 | 41.6 | fJ |
| SOT+FL Resistance* | $R_{SOT+FL}$ | 609 | 270 | 1385 | $\Omega$ |
| FinFET resistance | $R_{FET}$ | 5480 | 5260 | 5970 | $\Omega$ |

$\sigma_s$, $\lambda_s$ values for $\beta$-W, Pt, and Bi$_x$Se$_{(1-x)}$ are from references.
*Selected low write energy case @0.5ns

Compared with Pt and Bi$_x$Se$_{(1-x)}$, $\beta$-W's lower $I_{sw}$ gives rise to the lowest $E_{sw}^{FET}$, thus the best cell-level write energy performance. This illustrates again that as transistor resistance dominates the total cell resistance, optimizing for write current directly translates into lowest cell-level write energy.

## V. DISCUSSION

Last, we discuss how to expand our 2T-1MTJ modeling framework to study several variations of SOT-MRAM in addition to the type Y SOT-MRAM with a 2T-1MTJ cell above.

First, there are two additional types of SOT-MRAM: i.e., perpendicular SOT-MRAM (referred to as type Z) and in-plane SOT-MRAM with easy axis aligned in parallel to the current direction (referred to as type X). Note that the in-plane SOT-MRAM discussed above has an easy axis aligned orthogonal to the current direction (referred to as type Y).[25] As discussed in this work, if we only consider the effect of damping-like SOT, in type Z and type X devices, the damping-like SOT needs to overcome the magnetic anisotropy barrier without any precession, whereas, in type Y device, the damping-like SOT only needs to balance the damping constant, thus resulting in switching via multiple precessions.[36] Hence, the critical switching current of type Y is much lower than type X and type Z. Though experimentally, this is true,[25] only considering damping-like SOT is not sufficient for all three types of SOT-MRAM. For type Y, micromagnetic simulations indicate that the Oersted field might contribute to a much faster switching without incubation delay[37], which cannot be explained by the damping-like SOT-driven macrospin switching used in our model above. While for type X and type Z, experiments and macrospin simulations, show that field-like torque can assist the switching and lower the switching current.[25, 38] To account for these caveats mentioned above, micromagnetic simulations considering Oersted field, field-like SOT, and damping-like SOT are required.

Second, to enable high-density and low-cost MRAM for cache applications with sub-ns write performance, it is highly desirable to build a two-terminal SOT-MRAM with one single access transistor. Recently, one experimental work shows that, indeed, SOT can drive magnetization switching in a two-terminal MRAM cell consisting of a perpendicular MTJ sitting on top of a Ta SOT channel.[39] More detailed micromagnetic simulations are needed to fully understand the impact of non-uniform current distribution at the MTJ/SOT channel cross-section on SOT-driven switching. Besides, it is crucial to consider the effect of STT in this two-terminal scheme, which can lower the SOT switching current, as well as facilitate deterministic switching in type X and type Z devices.[38] We can employ the well-known STT-induced switching current equations.[24]

Third, utilizing the effect of voltage-controlled magnetic anisotropy (VCMA) can help temporarily lower the energy barrier during SOT switching, thereby further reducing the write current and energy.[40] We can model this VCMA-assisted SOT switching by including the VCMA-induced change of perpendicular magnetic anisotropy into the above model. Also, VCMA that lowers the energy barrier for both voltage polarities[41] can assist SOT switching in the two-terminal scheme mentioned above.

## VI. CONCLUSION

In this paper, we first introduce a 2T-1MTJ modeling framework, including practical transistor loading, SOT-induced switching, and current shunting of free layer over the SOT layer. The simplified framework shows that large spin Hall conductivity $\sigma_s$ and moderate sheet resistance $R_\square^{SOT}$ of 500 - 2000 $\Omega/\square$ are preferred for low switching current and energy. Using this framework, we benchmark the write current and energy performance of SOT-MRAM cells using SOT materials experimentally reported in the literature including heavy metals, topological insulators, and semimetals. A detailed benchmarking based on this framework further suggests $\beta$-W is a promising SOT material candidate for high-speed and low-power SOT-MRAM. We last discuss possible extensions beyond this 2T-1MTJ modeling framework for the modeling and design of various families of SOT-MRAM devices, circuits, and systems in the future.

ACKNOWLEDGMENT

The authors would like to thank Daniel Villamizar and Joseph Little for technical help.